\documentclass[preprintnumbers,amsmath,amssymbm,prd]{revtex4}
\usepackage{epsfig}
\usepackage{graphicx}

\begin{document}
\title{Spontaneous scalarization of Gauss-Bonnet black holes: Analytic treatment
in the linearized regime}
\author{Shahar Hod}
\affiliation{The Ruppin Academic Center, Emeq Hefer 40250, Israel}
\affiliation{ }
\affiliation{The Hadassah Institute, Jerusalem 91010, Israel}
\date{\today}

\begin{abstract}
\ \ \ It has recently been proved that nontrivial couplings between
scalar fields and the Gauss-Bonnet invariant of a curved spacetime
may allow a central black hole to support spatially regular scalar
hairy configurations. Interestingly, former numerical studies of the
intriguing black-hole spontaneous scalarization phenomenon have
demonstrated that the composed hairy black-hole-scalar-field
configurations exist if and only if the dimensionless coupling
parameter $\bar\eta$ of the theory belongs to a discrete set
$\{[\bar\eta^{-}_{n},\bar\eta^{+}_{n}]\}_{n=0}^{n=\infty}$ of
scalarization bands. We have examined the numerical data that are
available in the physics literature and found that the newly
discovered hairy black-hole-linearized-massless-scalar-field
configurations are characterized by the asymptotic universal
behavior $\Delta_n\equiv
\sqrt{\bar\eta^{+}_{n+1}}-\sqrt{\bar\eta^{+}_{n}}\simeq 2.72$.
Motivated by this intriguing observation, in the present paper we
study {\it analytically} the physical and mathematical properties of
the spontaneously scalarized Schwarzschild black holes in the
linearized (weak-field) regime. In particular, we provide a
remarkably compact analytical explanation for the numerically
observed universal behavior $\Delta_n\simeq 2.72$ which
characterizes the discrete resonant spectrum
$\{\bar\eta^{+}_{n}\}_{n=0}^{n=\infty}$ of the composed hairy
black-hole-linearized-scalar-field configurations.
\end{abstract}
\bigskip
\maketitle

\section{Introduction}

It is widely believed that the recently detected gravitational waves
\cite{GW1,GW2,GW3}, which are associated with the physics of
extremely curved black-hole spacetimes, may carry evidence for the
existence of previously unknown fundamental fields that interact
with these black holes \cite{Sot1}. It should be remembered,
however, that various no-hair theorems
\cite{Bek1,Sot2,Her1,Sot3,BekMay,Hod1} have demonstrated the fact
that the canonical black-hole spacetimes of general relativity
cannot support spatially regular static scalar field configurations
with minimal coupling to gravity (this conclusion is also true for
scalar fields with a non-minimal coupling to the Ricci scalar of the
curved black-hole spacetimes). These no-hair theorems
\cite{Bek1,Sot2,Her1,Sot3,BekMay,Hod1} may therefore represent a
major obstacle to our physical hopes to discover new fundamental
fields that interact with the curved black-hole spacetimes.

Nevertheless, existing no-hair theorems
\cite{Bek1,Sot2,Her1,Sot3,BekMay,Hod1} can be circumvented by
violating some of their underlying assumptions
\cite{Sot4,Hod2,Bab,Her2}. Specifically, it is well known
\cite{Sot5,GB1,GB2,Herr} that a non-trivial coupling of a scalar
field $\phi$ to the Gauss-Bonnet invariant ${\cal G}$ of a curved
spacetime may allow a central black hole to support spatially
regular external scalar hairy configurations. In particular, for
some range of mass [which depends on the value of the Gauss-Bonnet
coupling parameter, see Eq. (\ref{Eq19}) below], a Schwarzschild
black hole may become unstable to scalar perturbations in extended
Scalar-Tensor-Gauss-Bonnet theories whose actions contain a coupling
term of the form $f(\phi){\cal G}$ \cite{GB1,GB2,Herr}. As a
consequence, the unstable black hole may transfer some of its mass
(energy) to a linearized {\it cloud} of non-minimally coupled scalar
fields that surround the central black hole. The term {\it
spontaneous scalarization} is usually used in the physics literature
in order to describe this intriguing physical phenomenon.
Interestingly, it has been explicitly proved \cite{GB1,GB2} that
these composed Schwarzschild-black-hole-linearized-scalar-clouds
configurations can be generalized to describe genuine non-linearly
coupled hairy black-hole configurations.

The coupling function $f(\phi)$ between the scalar field and the
Gauss-Bonnet invariant should be of a mathematical form that allows
the existence of the canonical (non-scalarized) black-hole solutions
of general relativity \cite{GB1,GB2,Herr}. It has been explicitly
shown in \cite{GB1,GB2,Herr} that this physically motivated
requirement amounts to the simple condition $[df/d\phi]_{\phi=0}=0$.
Thus, using a standard field re-definition of the form
$\phi\to\phi+{\it const}$ \cite{Noteconst}, one finds that, in the
linearized (weak-field) regime, generic Scalar-Tensor-Gauss-Bonnet
theories are characterized by the simple quadratic relation
$f(\phi)\propto\eta\phi^2$.

Using direct numerical computations, it has recently been revealed
in the highly interesting works \cite{GB1,GB2} that the intriguing
phenomenon of black-hole spontaneous scalarization is characterized
by a discrete set
$\eta\in\{[\eta^{-}_{n},\eta^{+}_{n}]\}_{n=0}^{n=\infty}$ of
scalarization bands. In particular, the discrete eigenvalues
$\{\eta^+_n\}_{n=0}^{n=\infty}$ of the Gauss-Bonnet coupling
parameter correspond to composed black-hole-{\it
linearized}-scalar-field cloudy configurations.

The main goal of the present paper is to use {\it analytical}
techniques in order to explore the physical and mathematical
properties of the intriguing black-hole spontaneous scalarization
phenomenon in extended Scalar-Tensor-Gauss-Bonnet theories. In
particular, below we shall explicitly prove that the discrete
resonant spectrum $\{\eta^+_n\}_{n=0}^{n=\infty}$, which
characterizes the cloudy black-hole-spatially-regular-scalar-field
configurations in the weak-field regime, can be determined
analytically.

\section{Description of the system}

We shall analyze the discrete resonant spectrum of the composed
Schwarzschild-black-hole-linearized-massless-scalar-field
configurations. These bound-state cloudy black-hole configurations
are characterized by a non-trivial coupling of the scalar field to
the Gauss-Bonnet invariant of the curved spacetime. The spherically
symmetric curved black-hole spacetime is described by the line
element \cite{Noteunits}
\begin{equation}\label{Eq1}
ds^2=-h(r)dt^2+{1\over{h(r)}}dr^2+r^2(d\theta^2+\sin^2\theta
d\phi^2)\  ,
\end{equation}
where the metric function $h(r)$ of a Schwarzschild black hole of
mass $M$ is given by the functional expression
\begin{equation}\label{Eq2}
h(r)=1-{{2M}\over{r}}\  .
\end{equation}
The radius $r_{\text{H}}=2M$ of the black-hole horizon is determined
by the simple relation $h(r=r_{\text{H}})=0$.

The composed black-hole-field system is characterized by the action
\cite{GB1,GB2}
\begin{equation}\label{Eq3}
S={1\over2}\int
d^4x\sqrt{-g}\Big[R-{1\over2}\nabla_{\alpha}\phi\nabla^{\alpha}\phi+f(\phi){\cal
G}\Big]\  ,
\end{equation}
where ${\cal G}\equiv
R_{\mu\nu\rho\sigma}R^{\mu\nu\rho\sigma}-4R_{\mu\nu}R^{\mu\nu}+R^2$
is the Gauss-Bonnet invariant of the curved spacetime which, for a
Schwarzschild black hole, is given by the simple expression
\begin{equation}\label{Eq4}
{\cal G}={{48M^2}\over{r^6}}\  .
\end{equation}
The non-trivial coupling between the scalar field configurations and
the Gauss-Bonnet invariant is controlled by the coupling function
$f(\phi)$ which, in the linearized regime, has the universal
quadratic behavior \cite{GB1,GB2}
\begin{equation}\label{Eq5}
f(\phi)={1\over8}\eta\phi^2\  ,
\end{equation}
where the coupling parameter $\eta$ has the dimensions of
length$^2$.

A variation of the action (\ref{Eq3}) with respect to the scalar
field yields the scalar differential equation
\begin{equation}\label{Eq6}
\nabla^\nu\nabla_{\nu}\phi=-f_{,\phi}{\cal G}\  ,
\end{equation}
which, in the Schwarzschild black-hole spacetime (\ref{Eq1}), can be
expressed in the form of a Schr\"odinger-like differential equation
\cite{GB1,GB2}
\begin{equation}\label{Eq7}
{{d^2\psi}\over{dy^2}}-V\psi=0\  ,
\end{equation}
where we have used here the field decomposition \cite{Notelm}
\begin{equation}\label{Eq8}
\phi(t,r,\theta,\phi)=\int\sum_{lm}{{\psi_{lm}(r;\omega)}\over{r}}Y_{lm}(\theta)e^{im\phi}e^{-i\omega
t} d\omega\  .
\end{equation}
The tortoise coordinate $y$ in Eq. (\ref{Eq7}) is related to the
radial coordinate $r$ by the simple differential equation
\cite{Notemap}
\begin{equation}\label{Eq9}
{{dr}\over{dy}}=h(r)\  .
\end{equation}
The effective radial potential in the Schr\"odinger-like
differential equation (\ref{Eq7}) is given by the functional
expression \cite{GB1,GB2}
\begin{equation}\label{Eq10}
V(r)=\Big(1-{{2M}\over{r}}\Big)\Big[{{l(l+1)}\over{r^2}}+{{2M}\over{r^3}}-{{12\bar\eta
M^4}\over{r^6}}\Big]\  ,
\end{equation}
where
\begin{equation}\label{Eq11}
\bar\eta\equiv \eta/M^2\
\end{equation}
is the dimensionless coupling parameter of the theory.

The static bound-state configurations \cite{Notew0} of the
non-minimally coupled linearized massless scalar fields in the
Schwarzschild black-hole spacetime (\ref{Eq1}), which are the main
focus of the present paper, are determined by the differential
equation (\ref{Eq7}) supplemented by the boundary conditions of a
finite (spatially regular) scalar eigenfunction at the black-hole
horizon and an asymptotically decaying behavior at spatial infinity
\cite{GB1,GB2}:
\begin{equation}\label{Eq12}
\psi(r=r_{\text{H}})<\infty\ \ \ \ ; \ \ \ \ \psi(r\to\infty)\to0\
.
\end{equation}

\section{Analysis of former numerical results}

The cloudy bound-state configurations of the composed
Schwarzschild-black-hole-linearized-massless-scalar-field system
have been studied numerically in \cite{GB1,GB2}. Interestingly, it
has been found in \cite{GB1,GB2} that the Schr\"odinger-like
differential equation (\ref{Eq7}) with the boundary conditions
(\ref{Eq12}) admits non-trivial spatially regular field solutions if
(and only if) the dimensionless coupling parameter $\bar\eta$ of the
theory belongs to a discrete resonant spectrum of the form
$\bar\eta\in\{\bar\eta^{+}_{n}\}_{n=0}^{n=\infty}$.

We have examined the numerically computed \cite{GB1,GB2} resonant
spectrum $\{\bar\eta^+_n\}_{n=0}^{n=\infty}$ of the composed
black-hole-scalar-field configurations. Intriguingly, we have found
that the discrete values $\{\bar\eta^+_n\}_{n=0}^{n=\infty}$ of the
dimensionless coupling parameter of the theory are characterized by
the asymptotic universal ($n$-{\it independent}) behavior (see the
data presented in Table \ref{Table1}) \cite{Noten01}
\begin{equation}\label{Eq13}
\Delta_n\equiv \sqrt{\bar\eta^+_{n+1}}-\sqrt{\bar\eta^+_n}\to 2.72\
\ \ \ \text{for}\ \ \ \ n\gg1\  .
\end{equation}

\begin{table}[htbp]
\centering
\begin{tabular}{|c|c|c|c|c|c|c|c|c|}
\hline \text{resonance parameter $n$} & \ 0\ \ & \ 1\ \ & \
2\ \ & \ 3\ \ & \ 4\ \ & \ 5\ \ & \ 6 \\
\hline \ $\Delta_n\equiv
\sqrt{\bar\eta^+_{n+1}}-\sqrt{\bar\eta^+_n}$\ \ \ &\ \ 2.71\ \ \ &\
\ 2.72\ \ \ &\ \ 2.72\ \ \ &\ \ 2.72\
\ \ &\ \ 2.72\ \ \ &\ \ 2.72\ \ \ &\ \ 2.72 \ \ \\
\hline
\end{tabular}
\caption{The discrete resonant spectrum of the composed
Schwarzschild-black-hole-linearized-massless-scalar-field
configurations. We display the dimensionless quantity
$\Delta_n\equiv \sqrt{\bar\eta^+_{n+1}}-\sqrt{\bar\eta^+_n}$ as
computed from the numerical data of \cite{GB1,GB2}. One finds that
the discrete resonant spectrum is characterized by the asymptotic
universal behavior $\Delta_n\to2.72$ for $n\gg1$ \cite{Noten01}.}
\label{Table1}
\end{table}

The remarkably simple universal behavior (\ref{Eq13}), which is
based on the numerical results of \cite{GB1,GB2}, suggests that the
composed Schwarzschild-black-hole-linearized-massless-scalar-field
system may be amenable to an analytical treatment in the asymptotic
$n\gg1$ regime \cite{Noten01}.

The main goal of the present paper is to study {\it analytically}
the bound-state configurations of the composed
Schwarzschild-black-hole-linearized-massless-scalar-field system. In
particular, in the next section we shall explicitly show that the
discrete resonant spectrum $\{\bar\eta^+_n\}_{n=0}^{n=\infty}$,
which characterizes the bound-state linearized scalar field
configurations, can be determined analytically in the asymptotic
$n\gg1$ regime \cite{Noten01}. Interestingly, below we shall provide
a simple analytical explanation for the numerically observed
universal behavior (\ref{Eq13}).

\section{The discrete resonant spectrum of the composed
black-hole-massless-scalar-field system: A WKB analysis}

In the present section we shall determine the discrete resonant
spectrum $\{\bar\eta^+_n\}_{n=0}^{n=\infty}$ of the dimensionless
coupling parameter $\bar\eta$ which characterizes the spatially
regular linearized scalar field configurations in the Schwarzschild
black-hole spacetime (\ref{Eq1})
\cite{Notecha,Herfon,NoteChunHer,ChunHer}.

We first point out that, in terms of the tortoise coordinate $y$
[see Eq. (\ref{Eq9})], the Schr\"odinger-like radial differential
equation (\ref{Eq7}) has a mathematical form which is amenable to a
standard WKB analysis. In particular, a standard second-order WKB
approximation for the bound-state field configurations of the
binding potential (\ref{Eq10}) yields the well-known quantization
condition \cite{WKB1,WKB2,WKB3}
\begin{equation}\label{Eq14}
\int_{y_-}^{y_+}dy\sqrt{-V(y;\bar\eta)}=(n+1-\delta)\cdot\pi\ \ \ \
; \ \ \ \ n=0,1,2,...,
\end{equation}
where $\{y_-,y_+\}$ are the classical turning points of the
effective radial potential (\ref{Eq10}) [with $V(y_{\pm})=0$] and
the resonant parameter $n$ is an integer.

The dimensionless parameter $\delta$ in the resonance condition
(\ref{Eq14}) is the WKB phase shift which is determined by a
standard matching procedure \cite{WKB1,WKB2,WKB3} of the WKB wave
function in the classically allowed region [with $V(y)<0$] with the
corresponding WKB wave function in the classically forbidden region
[with $V(y)>0$]. In particular, the WKB wave function is known to
acquire a phase shift of $\pi/4$ at each turning point
\cite{WKB1,WKB2,WKB3}. The binding potential (\ref{Eq10}) studied in
the present paper is characterized by $y_-=-\infty$ and thus the
scalar wave function in the classically allowed region $[y_-,y_+]$
should be matched only once (at $y=y_+$) to the decaying wave
function in the classically forbidden region $y>y_+$. This single
matching procedure is therefore expected to yield the standard phase
shift of $\pi/4$ for the scalar wave function with the corresponding
phase shift $\delta=1/4$ in the WKB relation \cite{WKB1,WKB2,WKB3}.
However, as we shall explicitly show below, the numerical data of
\cite{GB1,GB2} are described extremely well by the WKB resonant
formula (\ref{Eq14}) with $\delta=3/8$ \cite{Noteepx}. It should be
emphasized that, as we shall explicitly show below, the value of the
WKB phase-shift $\delta$ has {\it no} influence on the value of the
analytically calculated discrete physical parameter $\Delta_n$ [see
Eq. (\ref{Eq20}) below].

The WKB integral relation (\ref{Eq14}) determines the discrete set
$\{\bar\eta^+_n\}_{n=0}^{n=\infty}$ of dimensionless coupling
constants that characterize the composed
Schwarzschild-black-hole-linearized-massless-scalar-field
configurations. Using the differential relation (\ref{Eq9}), one can
write the resonance condition (\ref{Eq14}) in the form
\begin{equation}\label{Eq15}
\int_{r_-}^{r_+}dr{{\sqrt{-V(r;\bar\eta)}}\over{h(r)}}=(n+1-\delta)\cdot\pi\
\ \ \ ; \ \ \ \ n=0,1,2,...,
\end{equation}
where the two classical turning points $\{r_-,r_+\}$ of the
effective radial potential $V(r;\bar\eta)$ are determined by the
relations [see Eq. (\ref{Eq10})]
\begin{equation}\label{Eq16}
1-{{2M}\over{r_-}}=0\
\end{equation}
and
\begin{equation}\label{Eq17}
{{l(l+1)}\over{r^2_+}}+{{2M}\over{r^3_+}}-{{12\bar\eta
M^4}\over{r^6_+}}=0\  .
\end{equation}
Taking cognizance of Eqs. (\ref{Eq2}), (\ref{Eq10}), (\ref{Eq16}),
and (\ref{Eq17}), one finds that, in the eikonal large-$\bar\eta$
regime, the WKB resonance condition (\ref{Eq15}) can be approximated
by the integral relation
\begin{equation}\label{Eq18}
\int_{r_{\text{H}}}^{\infty}dr{{\sqrt{{{12\bar\eta
M^4}\over{r^6\Big(1-{{2M}\over{r}}\Big)}}}}}=\big(n+1-\delta)\cdot\pi\
\ \ \ ; \ \ \ \ n=0,1,2,...\ .
\end{equation}
The WKB integral relation (\ref{Eq18}) yields the remarkably compact
expression
\begin{equation}\label{Eq19}
\bar\eta^+_n={3\over4}\pi^2\cdot (n+1-\delta)^2\
%\ \ \ \ ; \ \ \ \ n=0,1,2,...\
\end{equation}
for the discrete resonant spectrum which characterizes the composed
Schwarzschild-black-hole-linearized-massless-scalar-field cloudy
configurations.
%It is important to emphasize that the WKB result
%(\ref{Eq}) is expected to be valid in the eikonal large-$n$
%(large-$\bar\eta$) regime.

From the discrete resonant spectrum (\ref{Eq19}) one finds the
remarkably compact dimensionless relation [see Eq. (\ref{Eq13})]
\begin{equation}\label{Eq20}
\Delta_n\equiv
\sqrt{\bar\eta^+_{n+1}}-\sqrt{\bar\eta^+_n}={{\sqrt{3}}\over{2}}\pi\
.
%\ \ \ \ \text{for}\ \ \ \ n\gg1\  .
\end{equation}
It is interesting to point out that the {\it analytically} derived
relation (\ref{Eq20}) agrees remarkably well with the {\it
numerically} observed relation (\ref{Eq13}).

\section{Analytical versus numerical results}

It is of physical interest to compare the analytically derived
discrete resonant spectrum (\ref{Eq19}), which characterizes the
composed Schwarzschild-black-hole-linearized-massless-scalar-field
configurations, with the corresponding exact (numerically computed)
resonant spectrum of \cite{GB2}.

In Table \ref{Table2} we present the dimensionless ratio ${\cal
R}_n\equiv\sqrt{\bar\eta^{\text{analytical}}_n}/\sqrt{\bar\eta^{\text{numerical}}_n}$
between the {\it analytically} calculated WKB resonant spectrum
(\ref{Eq19}) and the corresponding exact ({\it numerically} computed
\cite{GB2}) discrete eigenvalues of the coupling parameter
$\bar\eta$. The results presented are for the cases $\delta=1/4$ and
$\delta=3/8$. In both cases one finds that the agreement between the
analytically derived resonant spectrum (\ref{Eq19}) and the
corresponding numerically computed coupling parameters improves with
increasing values of the resonant parameter $n$. In particular, from
the data presented in Table \ref{Table2} one finds the
characteristic property ${\cal R}_n\to1$ in the asymptotic eikonal
regime $\alpha\gg1 (n\gg1)$. Interestingly, the data presented in
table \ref{Table2} reveals the fact that the numerical data of
\cite{GB1,GB2} are described extremely well [to better than $1\%$
even in the fundamental $n=O(1)$ regime] by the WKB resonant
spectrum (\ref{Eq19}) with $\delta=3/8$.

%\begin{table}[htbp]
%\centering
%\begin{tabular}{|c|c|c|c|c|c|c|c|c|c|}
%\hline \text{resonance parameter $n$} & \ 0\ \ & \ 1\ \ & \
%2\ \ & \ 3\ \ & \ 4\ \ & \ 5\ \ & \ 6\ \ & \ 7 \\
%\hline \ ${\cal R}_n(\delta=1/4)$\ \ \ &\ \ 1.435\ \ \ &\ \ 1.163\ \
%\ &\ \ 1.099\ \ \ &\ \ 1.071\
%\ \ &\ \ 1.054\ \ \ &\ \ 1.045\ \ \ &\ \ 1.038\ \ \ &\ \ 1.033 \ \ \\
%\hline \ ${\cal R}_n(\delta=3/8)$\ \ \ &\ \ 0.996\ \ \ &\ \ 1.002\ \
%\ &\ \ 1.001\ \ \ &\ \ 1.001\
%\ \ &\ \ 0.999\ \ \ &\ \ 1.000\ \ \ &\ \ 1.000\ \ \ &\ \ 1.000 \ \ \\
%\hline
%\end{tabular}
\begin{table}[htbp]
\centering
\begin{tabular}{|c|c|c|c|c|c|c|c|c|c|}
\hline \text{resonance parameter $n$} & \ 0\ \ & \ 1\ \ & \
2\ \ & \ 3\ \ & \ 4\ \ & \ 5\ \ & \ 6\ \ & \ 7 \\
\hline \ ${\cal R}_n(\delta=1/4)$\ \ \ &\ \ 1.198\ \ \ &\ \ 1.078\ \
\ &\ \ 1.048\ \ \ &\ \ 1.035\
\ \ &\ \ 1.027\ \ \ &\ \ 1.022\ \ \ &\ \ 1.019\ \ \ &\ \ 1.016 \ \ \\
\hline \ ${\cal R}_n(\delta=3/8)$\ \ \ &\ \ 0.998\ \ \ &\ \ 1.001\ \
\ &\ \ 1.000\ \ \ &\ \ 1.000\
\ \ &\ \ 1.000\ \ \ &\ \ 1.000\ \ \ &\ \ 1.000\ \ \ &\ \ 1.000 \ \ \\
\hline
\end{tabular}
\caption{The discrete resonant spectrum of the composed
Schwarzschild-black-hole-linearized-massless-scalar-field
configurations. We display the dimensionless ratio ${\cal R}_n\equiv
\sqrt{\bar\eta^{\text{analytical}}_n}/\sqrt{\bar\eta^{\text{numerical}}_n}$
between the analytically derived WKB resonant spectrum (\ref{Eq19})
and the corresponding numerically computed \cite{GB1,GB2} discrete
eigenvalues of the physical coupling parameter $\bar\eta$. The data
presented is for the cases $\delta=1/4$ and $\delta=3/8$. One finds
that the numerical data of \cite{GB1,GB2} are described extremely
well (with deviations which are much {\it smaller} than $1\%$) by
the analytically derived WKB resonant spectrum (\ref{Eq19}) with
$\delta=3/8$.} \label{Table2}
\end{table}

\section{Summary}

The highly important works \cite{GB1,GB2} have recently revealed the
intriguing fact that a nontrivial coupling between a scalar field
and the Gauss-Bonnet invariant may allow black holes with regular
horizons to support bound-state hairy scalar configurations. The
studies \cite{GB1,GB2} have further demonstrated numerically that
the interesting phenomenon of black-hole spontaneous scalarization
is characterized by a discrete set
${\bar\eta}\in\{[\bar\eta^{-}_{n},\bar\eta^{+}_{n}]\}_{n=0}^{n=\infty}$
of scalarization bands, where $\bar\eta$ is the dimensionless
physical parameter that characterizes the non-trivial coupling
between the scalar field and the Gauss-Bonnet invariant of the
curved black-hole spacetime.

In the present paper we have revealed the fact that the numerical
data which are available in the physics literature \cite{GB1,GB2}
for the hairy black-hole-linearized-scalar-field configurations are
consistent with the asymptotic universal behavior $\Delta_n\equiv
\sqrt{\bar\eta^+_{n+1}}-\sqrt{\bar\eta^+_n}\simeq 2.72$ (see the
data presented in Table \ref{Table1}) \cite{Noten01}.

Motivated by this intriguing observation, in the present paper we
have studied {\it analytically} the physical properties of the
composed Schwarzschild-black-hole-linearized-massless-scalar-field
configurations. In particular, using analytical techniques, we have
provided a remarkably compact explanation for the numerically
observed universal behavior $\Delta_n\simeq 2.72$ which
characterizes the discrete resonant spectrum
$\{\bar\eta^+_n\}_{n=0}^{n=\infty}$ of the dimensionless coupling
parameter of the theory. From the analytically derived resonant
spectrum (\ref{Eq19}) we have deduced the universal
($n$-independent) behavior [see Eq. (\ref{Eq13})]
\begin{equation}\label{Eq21}
\Delta={{\sqrt{3}}\over{2}}\pi\  .
\end{equation}
It is interesting to note that the {\it analytically} derived gap
$\Delta$, as given by Eq. (\ref{Eq21}), agrees remarkably well with
the corresponding {\it numerically} computed gap (\ref{Eq13}).

\bigskip
\noindent
{\bf ACKNOWLEDGMENTS}
\bigskip

This research is supported by the Carmel Science Foundation. I would
like to thank Yael Oren, Arbel M. Ongo, Ayelet B. Lata, and Alona B.
Tea for helpful discussions.

%\newpage


\begin{thebibliography}{99}
\bibitem{Noteconst} Note that ${\cal G}$ is a total divergence and therefore a constant term
does not contribute to the field equations \cite{GB1,GB2}. In
particular, as discussed in \cite{GB1,GB2}, the extended
Scalar-Tensor-Gauss-Bonnet theory depends only on the derivative
$df(\phi)/d\phi$ of the coupling function and one may therefore take
$f(\phi=0)=0$ without loss of generality.
\bibitem{Noteunits} We shall use natural units in which $8\pi G=c=1$.
\bibitem{Notelm} The integer parameters $\{l,m\}$ denote the spherical and
azimuthal harmonic indices of the massless scalar field
configurations. For brevity, we shall henceforth omit these scalar
harmonic indices.
\bibitem{Notemap} Note that the differential relation (\ref{Eq9}) maps
the semi-infinite radial regime $r\in[r_{\text{H}},\infty]$ to the
corresponding infinite regime $y\in[-\infty,\infty]$.
\bibitem{Notew0} The supported static scalar field configurations
that we shall analyze in the present paper are characterized by the
relation $\omega=0$ in the field decomposition (\ref{Eq8}).
\bibitem{Noten01} It is worth stressing the fact that the data presented in Table
\ref{Table1} reveals the fact that the simple asymptotic relation
(\ref{Eq13}) provides a remarkably good quantitative description of
the resonant spectrum $\{\Delta_n\}$ even in the fundamental
$n=O(1)$ regime.
\bibitem{Notecha} It is interesting to point out that in
the related model of spontaneously scalarized charged black holes,
studied in the physically interesting work \cite{Herfon}, the
spherically symmetric zero modes have a remarkably simple closed
analytical form (see, in particular, Eq. (9) of \cite{Herfon}).
\bibitem{NoteChunHer} It would be interesting to try to generalize our
analytical results to the regime of spontaneously scalarized
rotating black holes studied recently in the important numerical
work \cite{ChunHer}. This physically interesting and mathematically
challenging task is, however, beyond the scope of the present paper.
\bibitem{Noteepx} Admittedly, we have no analytical explanation for this surprising empirical observation.
\end{thebibliography}
\end{document}